\documentclass[lnicst]{svmultln}
\usepackage{makeidx,graphicx,amssymb,units,subfigure}
\begin{document}
\mainmatter             
\title{Enhanced Free Space Beam Capture by Improved Optical Tapers}
\titlerunning{Enhanced Beam Capture by Improved Optical Tapers}  
%
\author{Tim Bartley\inst{1}$^,$\inst{2} \and Bettina Heim \inst{1}$^,$\inst{2} \and
Dominique Elser\inst{1}$^,$\inst{2} \and Denis Sych\inst{1}$^,$\inst{2} \and Metin Sabuncu\inst{1}$^,$\inst{2} \and Christoffer Wittmann\inst{1}$^,$\inst{2} \and Norbert Lindlein\inst{1}$^,$\inst{2} \and
Christoph Marquardt\inst{1}$^,$\inst{2} \and Gerd Leuchs\inst{1}$^,$\inst{2}}
\authorrunning{Tim Bartley et al.}   
%
\tocauthor{Tim Bartley, Bettina Heim, Dominique Elser, Denis Sych, Metin Sabuncu, Christoffer Wittman, Norbert Lindlein, Christoph Marquardt, Gerd Leuchs}
\institute{Institute of Optics, Information and Photonics, University of Erlangen-Nuremberg, Staudtstr. 7/B2, 91058 Erlangen, Germany\\
\and
Max Planck Institute for the Science of Light, G{\"u}nther-Scharowsky-Str. 1, Building 24, 91058 Erlangen, Germany\\
\email{t.bartley1@physics.ox.ac.uk}}

\maketitle              

\begin{abstract}        
In our continuous variable quantum key distribution (QKD) scheme, the homodyne detection set-up requires balancing the intensity of an incident beam between two photodiodes. Realistic lens systems are insufficient to provide a spatially stable focus in the presence of large spatial beam-jitter caused by atmospheric transmission. We therefore present an improved geometry for optical tapers which offer up to four times the angular tolerance of a lens. The effective area of a photodiode can thus be increased, without decreasing its bandwidth. This makes them suitable for use in our free space QKD experiment and in free space optical communication in general. 
\end{abstract}
\section{Introduction}\label{SECIntroduction}
Quantum Key Distribution (QKD) (reviewed in {e.g.}~\cite{Gisin02,Scarani08}) concerns the exchange of quantum states between two legitimate parties, conventionally named Alice and Bob. From these states, secret key data can be distilled. Unlike classical cryptography, for which the security is based on the unproven assumptions of computational difficulty, in principle QKD can be unconditionally secure. 


The first QKD protocol, BB84~\cite{Bennett84}, was demonstrated experimentally in 1992~\cite{Bennett92c}. Since then, numerous other protocols have emerged, {e.g.}~\cite{B92,Ekert91,Ralph99,Silberhorn02b}
. Such schemes have been established over very large distances, using both fibres (\unit[250]{km})~\cite{Stucki09} and free space (\unit[144]{km})~\cite{Schmitt-Manderbach07} as the quantum channel. 
In the absence of quantum repeaters~\cite{Briegel98}, free space optics (FSO) is required for worldwide quantum communication via satellites~\cite{Villoresi08,Bonato09}. Furthermore, FSO facilitates urban free space communication which would bypass the need and expense of new fibres being laid. This is one solution to the ``last mile'' problem currently faced by the telecommunications industry~\cite{Majumdar08}. 

Recently, we have demonstrated experimentally the feasibility of using continuous variables (CV), rather than single photons, to facilitate QKD~\cite{Lorenz04} in a real world free space environment with unrestrained daylight operation~\cite{Elser09}.
In this paper we explain why loss is of central importance to our scheme 
and we will present a new and practical approach to beam collection to reduce this particular loss. A characterisation of our set-up can be found in~\cite{Elser09}.




\subsection{Noise and Attenuation}
Noise and attenuation in the quantum channel 
are the limiting factors when determining the secure key rate. One of the central assumptions of any QKD protocol is that an adversary, Eve, has absolute control over the quantum channel. This means any and all attenuation in the channel are attributed to Eve, as well as any excess noise picked up during transmission.

In the continuous variable regime, attenuation gives Eve additional information and increases Bob's errors, 
both of which limit potential key rate~\cite{Heid06}. 
In principle, it is always possible to generate a secret key even under high losses~\cite{Silberhorn02}, although the rate becomes negligible if losses are too high. 

While we have shown that it appears no polarisation excess noise is present in the channel~\cite{Elser09}, it is worth considering the effects of intensity noise on quantum states~\cite{Dong08,Semenov09,Heim09}. Imperfect detection means security analysis is more involved and further post processing is required
. 
Security analysis under imperfect detection conditions, possibly caused by intensity noise, exists in the single photon regime~\cite{Fung09}. It remains to be seen what the implications are for continuous variables (we do not address these issues here), suffice to say that detection efficiency should be optimised as in all quantum information protocols, especially in the CV regime.

\section{Homodyne Detection}
\begin{figure}
\centering
\includegraphics[width=0.7\textwidth]{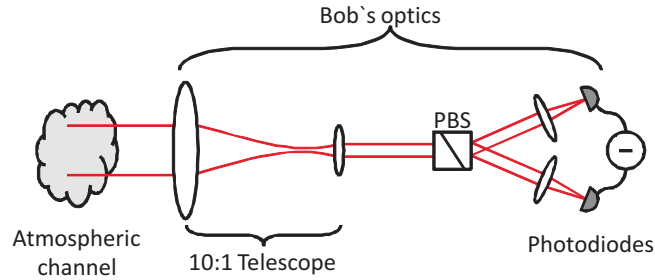}
\caption{Our free space homodyne detection. We use a Wollaston prism as the polarising beam splitter (PBS) in our set-up.}\label{DetSet}
\end{figure}

We measure the signal states using a balanced homodyne detection scheme, as shown in Fig.~\ref{DetSet}. Homodyne detection uses a local oscillator (LO) which interferes with the signal beam at a beam splitter. The difference of the two resulting photocurrents gives the amplitude of the signal state. Successful detection relies on splitting the incident intensity equally and amplifying the resulting photocurrent difference electronically (using an appropriate low-noise amplifier). This technique allows quantum noise to be measured using standard PIN photodiodes.

Conventionally, as the name implies, the local oscillator is generated locally by the receiver, {e.g.} in~\cite{Lange06}. However, using polarisation variables, we are able to multiplex the signal and LO in the same spatial mode at the sender. This results in perfect mode matching at Bob's beam splitter (as well as numerous other benefits explained in~\cite{Elser09}). However, since the LO is actually part of the detection system, the effect of loss is compounded. 

\subsection{Atmospheric Fluctuations}
In standard FSO, an intensity-modulated beam is focussed on a single diode~\cite{Majumdar08}. For a beam propagating from a source far away, all paths are considered paraxial, therefore only a lens is required to maximise detection efficiency. 

In our case, the homodyne detection set-up used to measure the polarisation states requires more optics. A telescope reduces the beam size before it is split at a polarising beam splitter (PBS) and focussed on the photodiodes, as shown in Fig.~\ref{DetSet}. These additional optics cause some paths of the beam to become non-paraxial in the presence of atmospheric beam jitter. This leads (in practice) to uncorrelated partial detection noise on each photodiode. However, any uncorrelated partial detection noise due to beam jitter leads to an imbalance across the diodes which may affect the homodyne detection~\cite{Elser09}. It is therefore desirable to remove the spatial dependence of detection due to a jittering beam.



\subsubsection{Compensating for Beam Jitter}
In principle, the active area of the photodiode can be increased, such that it captures the entire jittering beam. However, diode area scales with capacitance, which in turn limits bandwidth, reducing the speed of operation and thus key rate. 

Another strategy could be to focus using suitable lenses. However, imperfectly aligned aspheric lenses are susceptible to poor focussing of beams that are not incident normal to the surface to the lens ({i.e.} are not paraxial). 
Atmospheric beam jitter translates to angular deviations from the optical axis and the focus is no longer well defined. 
This effect is more pronounced in moving target implementations, such as surface to aircraft communications, {e.g.}~\cite{Horvath09}.
Hence the motivation to design an optical component which offers angular as well as spatial tolerance in its transmission behaviour for the receiver.

\section{Improved Optical Tapers}
Microscale optical tapers exist to couple beams between fibres and waveguides, see {e.g.~\cite{Burns86,Love91}}. 
They typically operate in the single-mode to single-mode regime. A similar implementation has been used for detectors in high energy physics~\cite{Hinterberger66,Winston70}, with ideas extending to solar power collectors using compound parabolic concentrators (CPCs)~\cite{Winston74} and generalised in terms of non-imaging optics~\cite{Winston05}.
While larger tapers have been suggested to operate in free space communication applications~\cite{Yun90}, to the best of our knowledge they have not been optimised and are not widely used. 
Using numerical ray trace analysis simulations, we present an improved geometry of an optical taper for free space communications purposes.

\subsection{Taper Geometry}
The aim of our taper is to collect all light incident on a large aperture and transmit it onto a photodiode of much smaller size. 
We want to effectively increase the active area of the photodiode without decreasing its speed of operation. 
Furthermore, we require the transmittance of the taper to be both spatially invariant with respect to the incident beam and offer higher angular tolerance than lenses. 

\subsubsection{Truncated Parabolic Mirror}
\begin{figure}
\centering
\begin{minipage}{0.45\textwidth}
\includegraphics[width=\textwidth]{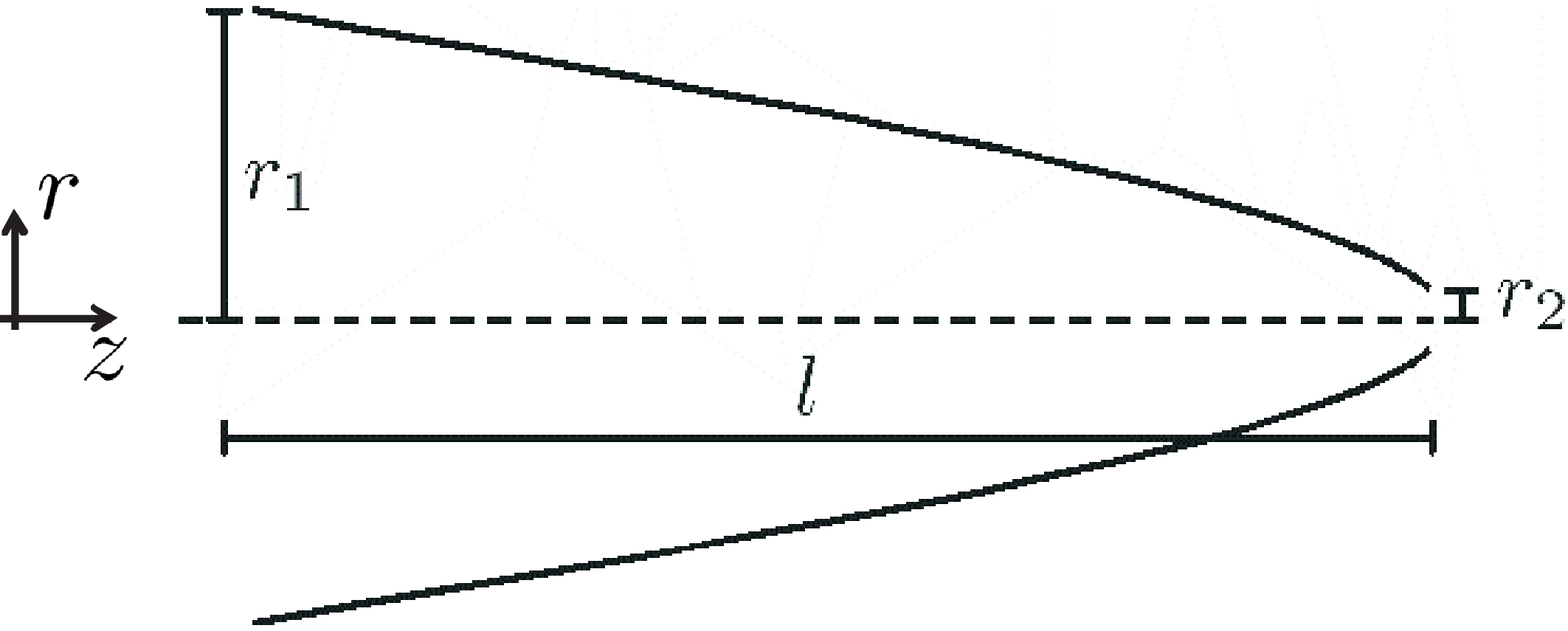}
\end{minipage}~\begin{minipage}{0.45\textwidth}
\includegraphics[width=\textwidth]{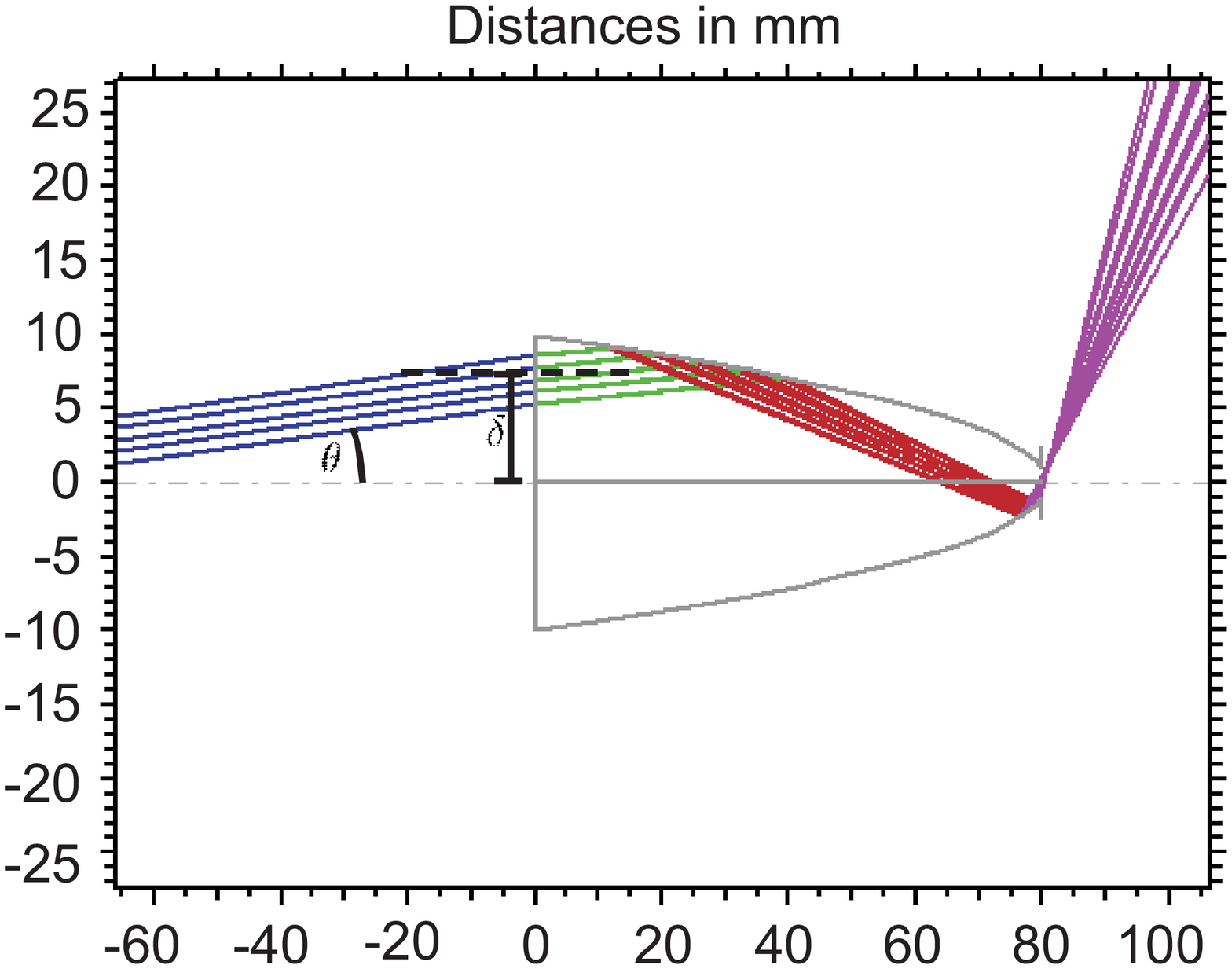}
\end{minipage}
\caption{Geometry and definitions of the parabolic taper (left). 
Example of ray trace of five rays (right). 
The colours of the rays change according to reflection/refraction events. 
Our actual simulations used over 10000 rays, from which transmission can be deduced by counting the resulting rays that pass through the aperture at the end of the taper. 
Software: RayTrace~\cite{RayTrace08}.}\label{Taper}
\end{figure}

One solution to the problem of compressing a wide incident beam to one point is a parabolic mirror. 
We therefore base our geometry on a parabolic fully-reflective surface, as shown in Fig.~\ref{Taper}~(left). 
The equation of a parabola is given by~$z\left(r\right)=\alpha r^2$, where~$r$ is the radial extension and $\alpha$ is a constant. 
In our case, the parabola is truncated in the~$z$-axis by the input and output apertures of radii~$r_1,r_2$ at~$z\left(r_1\right),z\left(r_2\right)$, respectively, such that the length of the taper is given by~$l=z\left(r_1\right)-z\left(r_2\right)$. 
The constant~$\alpha=\frac{l}{r_1^2-r_2^2}$ is thus written in terms of these parameters.

We require all the incident light to exit the taper, {i.e.} the focus $z_f$ of the parabola lying outside the taper ($z_f>z\left(r_2\right)$). This imposes the condition that the gradient $r^\prime\left(z\right)$ of the parabola (with respect to the $z$-axis) can never exceed unity, otherwise some paths within the taper would be back-reflecting. If we assume~$r_1$ and~$r_2$ are fixed by the size of beam jitter and diode area respectively, we therefore seek a taper of length~$l$ such that the condition~$\left|r^\prime\left(z\right)\right|<1$ is fulfilled. Given the gradient~$r^\prime\left(z\right)=\frac{1}{2\alpha r}$ is maximal at~$r=r_2$, upon substituting for~$\alpha$ the condition above becomes 
\begin{equation}
\left|\frac{r^2_1-r_2^2}{2lr_2}\right|<1~.
\end{equation}
This means that the length~$l$ is limited by
\begin{equation}
l>\frac{\left(\beta^2-1\right)r_2}{2}~,
\end{equation}
where~$\beta=\frac{r_1}{r_2}$ is the ratio of the input and output aperture radii of the taper.


\subsection{Numerical Simulation}
It is important to stress that unlike conventional tapers ({e.g.}~\cite{Burns86,Love91}), we are operating in the highly multimode regime, rather than guiding single mode to single mode. 
The large size of the taper compared to the wavelength of the light permits the use of ray approximations rather than wave optics. 
Using the in-house numerical analysis programme RayTrace~\cite{RayTrace08}, the response of the taper to geometric rays can be simulated. 
An example of a trace is shown in Fig.~\ref{Taper}~(right). Each incident ray represents an amount of energy governed by a Gaussian distribution centred on the middle ray. 
The incident rays can be parameterised by a radial and angular displacement vector $\left(\delta,\theta\right)$, as used in ray transfer matrices. 
Using this technique, the transmission of the taper for different values of $\delta$ and $\theta$ can be calculated. 


\begin{figure}
\centering
\includegraphics[width=\textwidth]{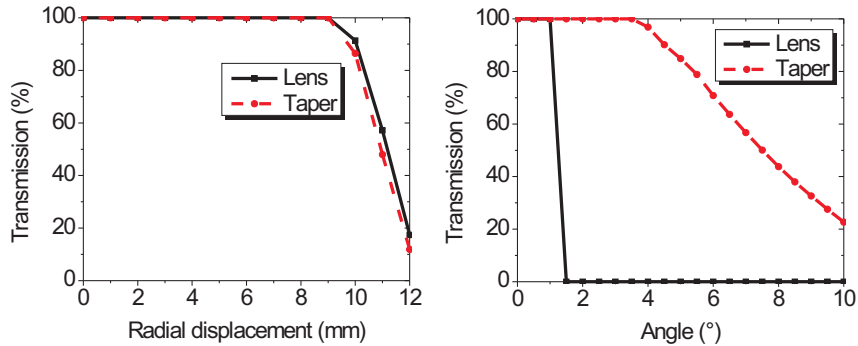}
\caption{Taper and lens transmission as a function of radial displacement (left) and angle (right). 
Both components show similar radial dependence. 
However, the lens is only fully transmittant up to angular deviations of about 1$^\circ$, whereas the taper transmits fully up to about 4$^\circ$. 
}\label{Trans}
\end{figure}

We simulate a beam of width \unit[2]{mm} (similar to beam widths used in our QKD set-up) incident on a taper with fully-reflecting side facets of initial aperture \unit[20]{mm} and final aperture of \unit[2]{mm}, as shown in Fig.~\ref{Taper}~(right). 
An absorbing plane with a \unit[2]{mm} aperture is placed immediately after the taper. 
The transmission through this aperture represents the total intensity that would impinge on a photodiode of diameter \unit[2]{mm}. 
Figure~\ref{Trans} (left) shows the transmission as a function of radial displacement~$\delta$ (with~$\theta=0~\forall~\delta$) for a taper compared to a lens of equal aperture. 
The ``photodiode'' is placed at the focus of the lens. As can be seen, both optical components serve to increase the effective diameter of the aperture to \unit[17]{mm}.


The angular dependence is tested slightly differently. 
For a taper of non-linear geometry, whether or not a ray will be transmitted depends on both angle of incidence and point of entry into the taper. 
We therefore simulate a number of beams of equal amplitude spread across the input aperture. 
We test this at a number of different angles, with the results shown in Fig.~\ref{Trans}~(right). 
The lens is only fully transmittant up to 1$^\circ$, whereas the taper offers four times more angular tolerance, remaining fully transmittant up to 4$^\circ$ and offering at least \unit[50]{\%} efficiency up to 7$^\circ$. 

\begin{figure}
\centering
\includegraphics[width=\textwidth]{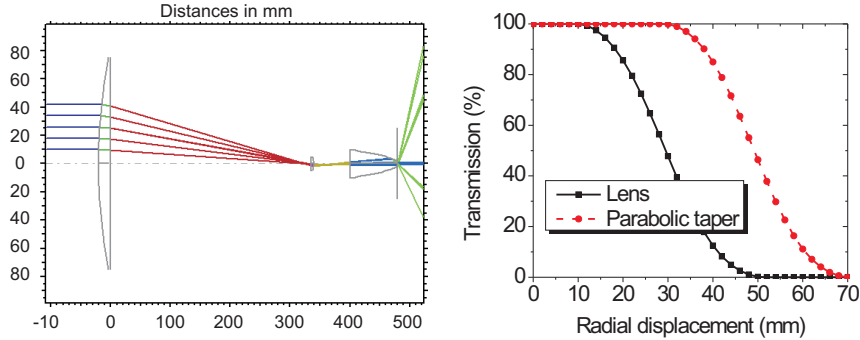}
\caption{Left: ray trace of the detection set-up including telescope. We see how angular deviations build up due to the 
optics. Right: transmission to a photodiode as a function of radial displacement at the initial telescope. The taper offers tolerance up to \unit[30]{mm}, amounting to a three-fold improvement over a lens when coping with beam jitter.}\label{TelTrace}
\end{figure}

We also simulated how these general results translate to improvements in our specific set-up~\cite{Elser09}. The receiver telescope and taper are shown in the ray trace in Fig.~\ref{TelTrace} (left). Here, we vary the displacement incident on the receiver telescope, as would be the case for a jittering beam. For stationary sender and receiver stations, we do not need to consider angular variations since the beam propagating from a source far away is considered paraxial. However, radial displacement here is converted to angular deviations on the way to the photodiode, 
as shown in Fig.~\ref{TelTrace} (left). These angular deviations arise from the realistic case of slight misalignment of the telescope or lens aberrations. Explicitly quantifying the tolerance of axial misalignment of the telescope is difficult. Here we simulate lens positions within \unit[1]{mm} of perfect alignment, which could be reasonably expected experimentally in our set-up. 
The improvement of the taper over a lens is shown in Fig.~\ref{TelTrace} (right). In our set-up, the taper offers considerably more tolerance than a lens, remaining fully transmittant up to radial deviations at the telescope of up to \unit[30]{mm}. 

\section{Outlook}
We have begun characterising such tapers with respect to our quantum communication experiment. 
Their performance for a jittering beam can be analysed in our existing \unit[100]{m} free space channel~\cite{Elser09}. 
We then plan to implement such tapers in our \unit[1.6]{km} point-to-point QKD link, which is currently under construction.

Conventional tapers are made out of glass and rely on total internal reflection to guide light. 
In our case, the rays may strike the glass/air boundary outside the range of the critical angle, such that some portion would be lost. 
This leads to the undesired effect of position-dependent transmission. 
To counter this, we coat our glass tapers with silver (offering high reflectivity for our wavelength of \unit[800]{nm}), negating the critical angle condition for reflection. 
We also plan to coat the input and output apertures with an anti-reflective layer for our wavelength.

At the moment we study glass cores derived from the waste products of a fibre-drawing machine. 
As a result, they are currently not machined precisely to our specifications. 
However, constructing arbitrary taper geometries is possible in principle \cite{Birks92}. 
An alternative approach would be to use a hollow taper with a highly-reflective inner surface.

\subsubsection{Acknowledgements} The authors would like to thank Silke Rammler and Leyun Zang for useful discussions, and Paul Lett for correspondence.

%
%

%
\end{document}